\documentstyle[prc,preprint,aps,epsf,rotate]{revtex}
\begin{document}
\tighten
\title{Muon capture on nuclei with $N > Z$, 
random phase approximation, and
in-medium renormalization of the axial-vector coupling constant}
\author{E. Kolbe}
\address {Physics Department,
University of Basel, CH-4056 Basel, Switzerland} 
\author{K. Langanke}
\address{Institute for Physics and Astronomy,
University of Aarhus, DK-8000 Aarhus, Denmark}
\author{ P. Vogel}
\address{Physics Department\\
California Institute of Technology, Pasadena, California 91125 USA}
\date{\today}
\maketitle

\begin{abstract}
We use the  random phase approximation to describe the muon
capture rate on ${}^{44}$Ca,${}^{48}$Ca,
${}^{56}$Fe, ${}^{90}$Zr, and ${}^{208}$Pb. 
With ${}^{40}$Ca as a test case,
we show that the Continuum Random Phase Approximation (CRPA)
and the standard RPA give essentially equivalent descriptions
of the muon capture process. Using the standard RPA
with the free nucleon weak form factors we 
reproduce the
experimental total capture rates on these nuclei quite well. 
Confirming our previous CRPA result for the
$N = Z$ nuclei, we find that the calculated rates would be
significantly lower than the data if the in-medium
quenching of the axial-vector coupling constant were employed.
\end{abstract}

The capture of a negative muon from the atomic $1s$ orbit,
\begin{equation}
\mu^- + (Z,N) \rightarrow \nu_{\mu} + (Z-1, N+1)^*
\end{equation}
is a semileptonic weak process which has been studied for a long time
(see, e.g., the reviews by Walecka \cite{Wal} or Mukhopadhyay
\cite{Muk} and the earlier references therein). The total capture
rate has been measured for many nuclei \cite{Su87}; in some cases the partial
capture rates to specific states in the daughter nucleus have been
determined as well.

The nuclear response in muon capture is governed by the momentum
transfer which is of the order of the muon mass. The energy transferred
to the nucleus is restricted from below by the mass difference of the
initial and final nuclei, and from above by the muon mass. The phase
space and the nuclear response favor lower nuclear excitation
energies. There is an intimate relation between the inclusive
muon capture rate and the cross section for the antineutrino-induced
charged-current reactions; both are governed by the same
nuclear matrix elements and proceed from the same initial to 
the same final states. 

Since the experimental data are quite precise, and the theoretical
techniques of evaluating the nuclear response in the relevant regime
are well developed, it is worthwhile to see to what extent the
capture rates are theoretically understood. This is not
only interesting per se, but should be viewed
as a more general test of our ability
to describe semileptonic weak charged-current reactions.

A step in this direction was
undertaken by us several years ago in Ref. \cite{us} where 
the continuum random phase approximation 
(CRPA) was used to describe  muon
capture on the $N=Z$ nuclei
${}^{12}$C, ${}^{16}$O and ${}^{40}$Ca. We showed that the
method allows us to reproduce the
experimental total capture rates on these nuclei to better than 10\%
using the free nucleon weak form factors and two different residual
interactions. In particular, it was not necessary
to apply the in-medium quenching of the axial vector coupling constant.
This is contrary to various well-known indications that
the axial-vector coupling constant $g_A$ in nuclear medium is reduced
from its free nucleon value of $g_A=1.26$ to the value of
$g_A\simeq1$ when one analyses
the data on  beta decay between low-lying states of the
$(sd)$ shell nuclei \cite{Wild} and $(pf)$ shell
nuclei \cite{Zuker}. In addition, the ``missing
Gamow-Teller strength'' problem, as revealed in the interpretation of
the forward-angle $(p,n)$ and $(n,p)$ 
charge-exchange reactions \cite{pn}, is
also often quoted as evidence for quenching of $g_A$. 
Note that the
Gamow-Teller (GT) strength is concentrated in the giant GT resonance at
excitation energies not very far from the energies involved in the
muon capture, although this latter process is usually dominated by the
transitions to the negative parity spin-dipole states. 
That is so in particular in the double-magic nuclei  
${}^{16}$O and ${}^{40}$Ca where the GT strength is strongly suppressed,
while in $^{12}$C it is essentially exhausted by the transition to
the analog of the $T=1$ state at 15.11 MeV in $^{12}$C, whose contribution
to the muon capture rate has been subtracted in \cite{us}. 
It is thus of
interest to inquire whether a similar quenching applies in muon capture
over a broader range of nuclei.
In this way, we can test whether the quenching is
a general phenomenon, applicable to more than just GT transitions. 

In this paper we extend the previous calculation \cite{us} to heavier
nuclei, in particular nuclei with the neutron excess, i.e., with
a nonvanishing value of the initial isospin. 
In \cite{us} the Continuum Random Phase Approximation (CRPA)
was used, a method shown to be
successful for the description of the nuclear response to weak and
electromagnetic probes \cite{RPA}. The method combines the usual RPA
treatment with the correct description of the continuum nucleon decay
channel. For heavier $N > Z$ nuclei that method is computationally
quite demanding. Moreover, it can describe  only transitions
to states in the daughter nucleus above the particle emission
threshold. In order to evaluate the total capture rate
one has to include also the transitions to bound states, which
contribute relatively more, and are typically not experimentally
separated in the heavier nuclei.

\begin{figure}[htb]
\begin{center}
{\epsfxsize=0.8\textwidth
\epsffile{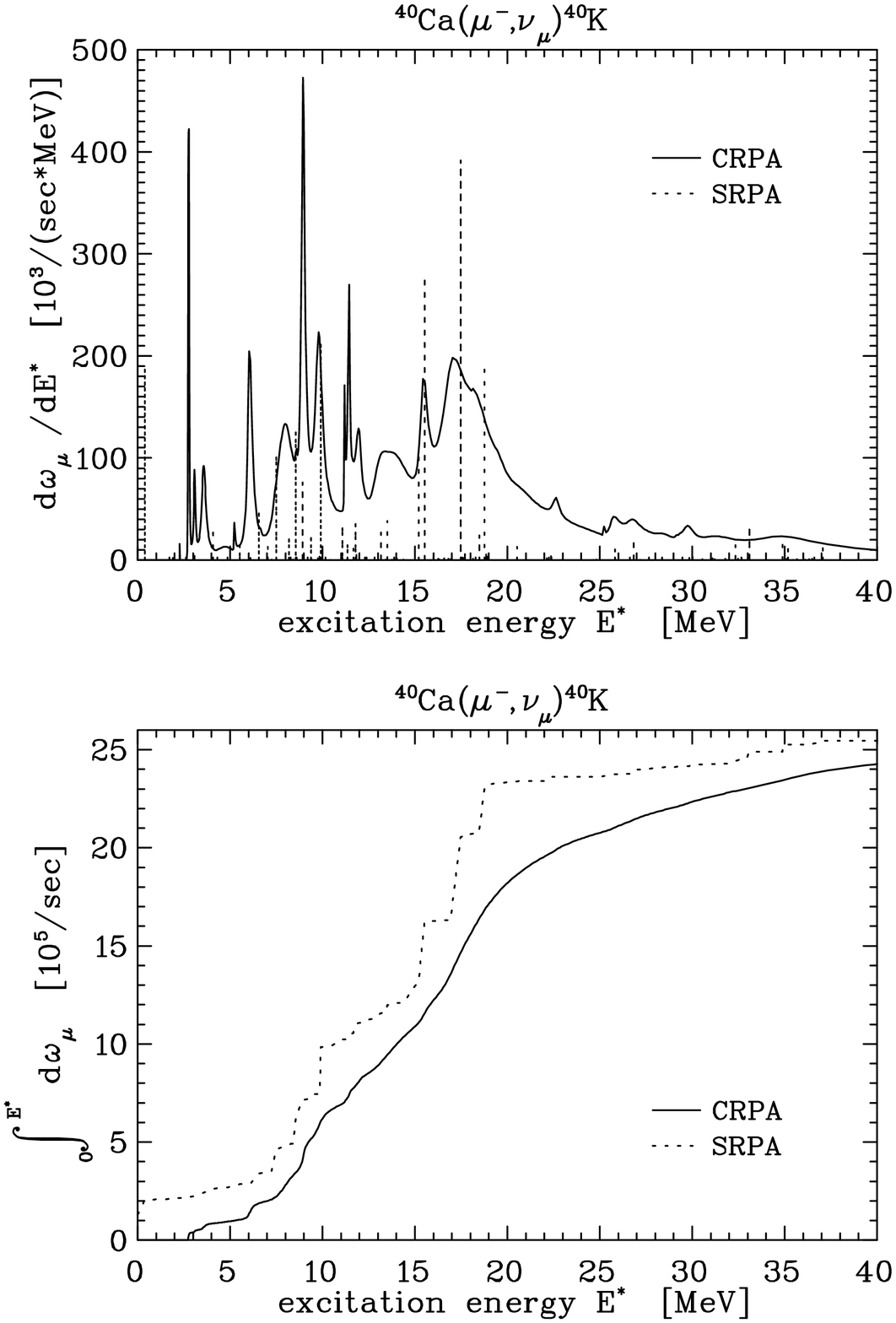}}
\caption{The $\mu^-$ capture rate as a function of the excitation energy
in the final nucleus $^{40}$K (upper panel). 
The continuous curve is for the continuum
random phase approximation (CRPA), while the dashed vertical bars are 
the results of the standard random phase approximation (SRPA).
In the lower panel the integrated $\mu^-$ capture rate, 
up to the excitation energy $E^*$, is shown.}
\label{fig:1}
\end{center}
\end{figure}

If one is interested in the {\it total} capture rate, the numerically
simpler standard RPA is just as good. As an additional bonus, it
avoids the distinction between the bound and unbound states.
That the two methods, CRPA and standard RPA, are equivalent
for the present purpose is demonstrated in Fig. \ref{fig:1}
for the case of $^{40}$Ca. We show in the upper part
the differential capture rate as a function of the excitation
energy in the final nucleus. While the CRPA (for details
see Refs. \cite{RPA}) is characterized by 
a continuous curve, nonvanishing everywhere above the threshold,
the standard RPA is characterized by the `picket fence', since
there is a finite number of discrete final states. The similarity
of both methods is even better seen in 
the lower part of Fig. \ref{fig:1} which shows the
integrated rate up to a given excitation energy $E^*$. 
There appears to
be a slight systematic shift of a few MeV (caused by the bound state
contribution, presumably), but the final
capture rates, and the typical excitation energies, are
remarkably similar. Thus, we use the standard RPA for the
evaluation of the muon capture in the selected $N > Z$ nuclei.

For the calculation in the present work we used the
phenomenological Landau-Migdal force with parameters that has
been shown to be applicable for a wide range of nuclei
\cite{LM}. All single-particle states below the Fermi
level were included and two oscillator shells above
it were taken into account. Using again
$^{40}$Ca as a test case, we checked that adding or subtracting
in the calculation
few subshells above the Fermi level does not visibly change the muon
capture rate. However, in $^{208}$Pb enlarging the
single particle space leads to an increase of
the capture rate by about 5\%. The free nucleon
form factors were used to describe the weak nuclear current.
In particular, the unquenched axial vector coupling constant
$g_A (0) = 1.26$ was used. 

Muon capture also depends on the induced pseudoscalar hadronic weak
current. At the free nucleon level the corresponding coupling
constant is determined by the Goldberger-Treiman relation \cite{Gold}
\begin{equation}
F_P(q^2) = \frac{2 M_p g_A (0)} {m_{\pi}^2 - q^2}\;,
\end{equation}
where $m_{\pi}$ is the pion mass and $g_A(0) = 1.26$. (In
muon capture one often uses a dimensionless quantity $g_P=m_{\mu}
F_P(q^2)$ at the relevant momentum transfer $q^2\simeq
-0.9m_{\mu}^2$, such that $g_P\simeq8.4$ for free protons.) In
nuclear medium $F_P$ can be again renormalized, and this
renormalization does not necessarily obey the Goldberger-Treiman
relation \cite{Magda}. We have shown in our previous work that 
the total muon capture rates are not sensitive enough to the
various choices of $F_P$ renormalization. Consequently, throughout
this work we use the Goldberger-Treiman relation. 

The calculated total capture rates are collected in Table \ref{tab:1}
and compared with the data \cite{Su87}. For comparison 
we also show in Tab. \ref{tab:1} the
earlier results  \cite{us} for the $N = Z$ nuclei, evaluated
with the same residual Landau-Migdal force.
Among the nuclei in Table \ref{tab:1}, $^{56}$Fe is the only one with
a substantial contribution of the Gamow-Teller transitions,
which are known to be quenched. Hence the $1^+$ part of the
capture rate was quenched by the empirical factor of 2.54
obtained by comparing the total experimental and calculated
GT strength. This reduced $1^+$ rate was then combined with
the unquenched rate from the other multipoles in column 3
of Table \ref{tab:1}.
Clearly, a  better agreement with experiment
for all considered nuclei is achieved
when the full value of $g_A$ is used. The quenched value leads
to an obvious underestimate of the muon capture rate.

\begin{figure}[htb]
\begin{center}
\epsfxsize=5.25in \epsfbox{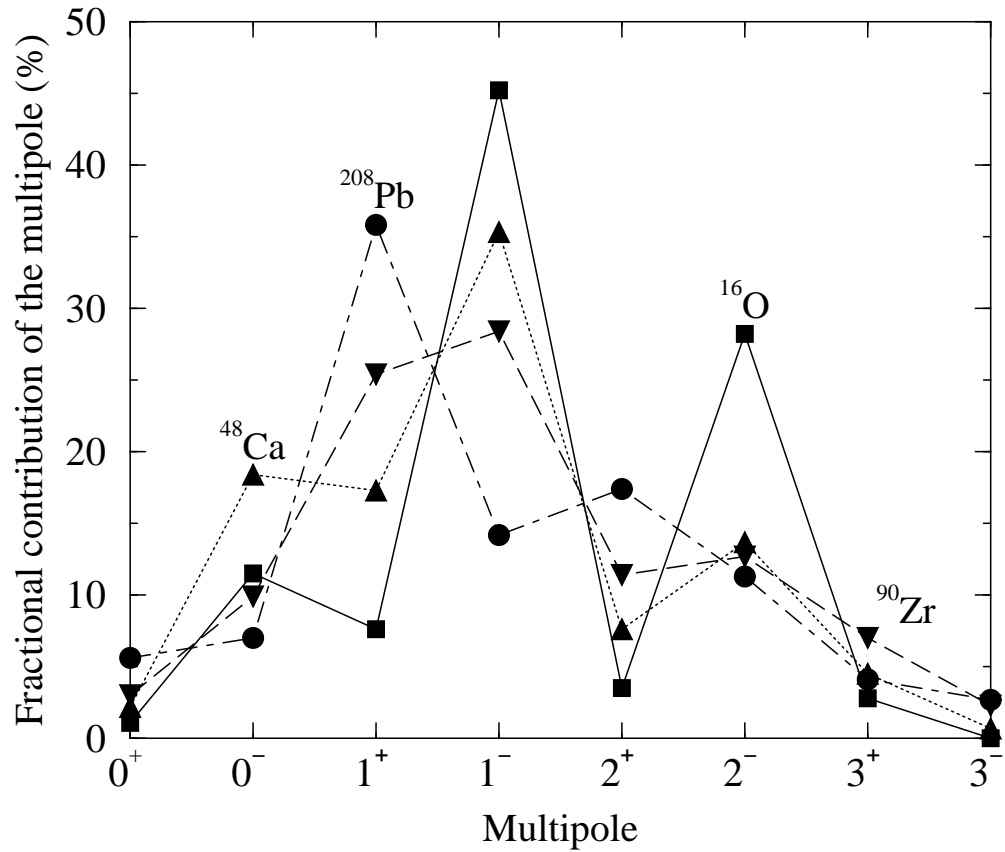}
\caption{Fractional contribution of different multipoles
to the total capture rate for four of the considered nuclei.
The entries with $I^{\pi} = 0^-,1^-, 2^-$ correspond to the
1$\hbar\omega$ excitations, while the entries with
$I^{\pi} = 1^+,2^+, 3^+$ correspond to the
2$\hbar\omega$ excitations. The data points are connected
by lines for better visibility and the nuclei are as indicated in the 
figure.}
\label{fig:3}
\end{center}
\end{figure}

In Fig. \ref{fig:3} the fractional contributions of different
multipoles are shown for $^{16}$O, $^{48}$Ca, $^{90}$Zr,
and $^{208}$Pb. For the closed shell nucleus  $^{16}$O
the negative parity $1^-$ and $2^-$, 
and to lesser extent $0^-$, multipoles dominate, as expected.
In the intermediate mass nuclei, $^{48}$Ca and $^{90}$Zr,
the dominance of these multipolarities is less pronounced.
Finally, in $^{208}$Pb the positive parity $1^+$ and $2^+$ multipoles
give the largest contribution, since the negative parity
proton hole-neutron particle states are blocked. Note, however,
that these $1^+$ and $2^+$ states correspond to the 2$\hbar\omega$
excitations involving different major shells, not the 
usual 0$\hbar\omega$ Gamow-Teller transitions.

The dominance of dipole transitions 
for the $\mu^-$ capture process in the $p$ and $s,d$ shell nuclei
is a well known phenomenon. It can be exploited in the
generalized Goldhaber-Teller model \cite{DDH}, where all dipole
and spin-dipole strength is concentrated in a single collective
state. The  Goldhaber-Teller model has been applied to the muon
capture already a long time ago \cite{DDH}. Here we have repeated the
calculation with one modification. Instead of using as
the energy of the single collective state the energy 
$E_D$ of the giant dipole resonance in the initial nucleus,
we placed the strength into a state at the
excitation energy $E^* = E_D - E_{IAS}$ in the final nucleus,
where $E_{IAS}$ is the energy of the isobar analog state 
with $T=1$ in the initial $N = Z$ nucleus. Such an assignment
puts the strength close to the centroid of the excitation spectrum
obtained in the RPA. The method is rather crude, since
more detailed calculations clearly show that the spin-dipole
strength is spread over a sizable energy interval. Nevertheless,
the results displayed in Table \ref{tab:2}, obtained again with
the full value of $g_A(0) = 1.26$, support our conclusion
that there is no quenching in these nuclei of the operators 
of the weak current that change parity. 

First forbidden beta decays, 
in particular those with $\Delta I^{\pi} = 0^-$,
have been often analyzed as a source of information on the 
enhancement caused by the meson exchange currents. 
In the context of the work reported here it is worthwhile to quote
the work of Warburton and Towner \cite{War} who analyzed 18
first forbidden beta decays in the lead region. They used a truncated
shell model and found that the  $\Delta I^{\pi} = 1^-$ transitions,
which are unaffected by the meson exchange currents, do not
require any quenching. In fact, their fit to an overall quenching factor
for the first forbidden matrix elements
results in $sq_1 = 0.98 \pm 0.05$, compatible with unity, 
i.e., with no quenching.
That analysis, totally independent of our evaluation of the
muon capture, again supports the conclusion about the apparent
absence of an appreciable quenching of the parity-changing
(i.e., first forbidden) weak current operators.

The unique second forbidden beta decays are governed by a single
operator $r^2[Y_2\sigma]^{\lambda=3}$. For only a handful of them 
the partial decay rates are known. When analyzed \cite{Gabr98,Warb92},
these transitions do not allow one to draw any definitive conclusion
about the possible quenching of the corresponding strength.
However, as pointed out above, $\mu^-$ capture rate in $^{90}$Zr,
and  $^{208}$Pb, is strongly affected by the 
2$\hbar\omega$ transitions, whose operators are related to the
second forbidden beta decays. Since the agreement between
the experimental and calculated rates in these two nuclei,
and in particular in $^{208}$Pb, is 
not as good as in the other cases, we cannot make a strong
statement regarding the quenching of the positive parity
``second forbidden'' multipoles based on the muon capture 
calculations.

In conclusion, the present analysis shows that the CRPA and SRPA
methods are capable of describing the total $\mu^-$ capture
rates quite well in a large range of nuclei. 
The dependence of the muon capture rate
on the isospin, the so-called Primakoff rule \cite{Prim},
is also reasonably well reproduced. There is no indication
of the necessity to apply any quenching to the
operators responsible for the $\mu^-$ process.
Thus our findings indicate that any in-medium quenching of the
axial current matrix elements appears to be restricted to the
0$\hbar\omega$  spin changing
operators, i.e. to the Gamow-Teller operator.

\acknowledgements
This work was
supported in part by the Swiss National Science Foundation,
the Danish Research Council,
and by the U.S. Department of Energy, Contract
\#DE-F603-88ER-40397.

\begin{table}[htb]
 \caption[]{$\mu^-$-capture rates calculated within the standard 
            (SRPA) and 
            continuum (CRPA) models in units of $10^3/s$. 
            The radius $r$ and diffuseness $d$ of 
            the extended nuclear charge distribution were set to 
            $(r,d)$ = (1.07 fm,0.50 fm) for $^{12}$C and
            $(r,d)$ = (1.07 fm,0.57 fm) for all other nuclei.
            The Landau-Migdal force (LM) is used throughout.}
 \protect\vspace{1ex}
 \label{tab:1}
 \begin{center}
 \begin{tabular}{|c||c|cc|cc|} \hline
   nucleus    & Exp.~\cite{Su87} & SRPA  & SRPA  & CRPA(LM) & CRPA(LM) \\
             & & $g_A(0)=1.26$ & $g_A(0)=1.0$ & $g_A(0)=1.26$ & $g_A(0)=1.0$
    \\[1ex] \hline \hline
   $^{\rm 12}$C & $32.8\pm0.8$   &          &          & 31.3$^{(3)}$
 &  22.9$^{(3)}$  
    \\[2ex] \hline 
   $^{16}$O   & $102.6\pm0.6$    &          &          & 103.2 & $ 75.8 $ 
    \\[2ex] \hline  
   $^{40}$Ca  & $2544\pm 7^{(1)}$        &    2547  &    1846  &   2489 & 1800 
    \\[2ex] \hline 
   $^{44}$Ca  & $ 1793\pm 40 $   &    1722  &    1238  &        & 
    \\[2ex] \hline 
   $^{48}$Ca  &  1164$^{(2)}$      &    1301  &     930  &        & 
    \\[2ex] \hline
   $^{56}$Fe  & $ 4400\pm 100 $  &    4460$^{(3)}$  & 
   3430$^{(3)}$  &        & 
     \\[2ex] \hline 
   $^{90}$Zr  & $ 9350\pm 100 $  &   10288  &    7400  &        & 
    \\[2ex] \hline 
   $^{208}$Pb & $ 13450\pm180 $  &   16057  &   11436  &        & 
    \\[2ex] \hline  
 \end{tabular}

\medskip

$^{(1)}$ Corrected from the data for natural Ca \\
$^{(2)}$ Extrapolated using the Primakoff 
formula fitted to $^{40}$Ca and $^{44}$Ca \\
$^{(3)}$ Calculated with partial occupation of the single particle 
subshells, see {\protect \cite{c12new}}.
\end{center}
\end{table}

\narrowtext
\begin{table}[htb]
\caption[]{$\mu^-$-capture rates calculated within the 
Goldhaber-Teller model in units of $10^3$/s. 
The formulae of Ref. {\protect \cite{DDH}} are used.
In column 2 the original parameters are employed. In column
3 the excitation energy $E^*$ is modified, as explained in the text.}
 \protect\vspace{1ex}
 \label{tab:2}
\begin{center}
\begin{tabular}{|c||c|c|c|} \hline
   nucleus    & Exp.~\cite{Su87} &  orig. values {\protect \cite{DDH}} 
 &  modified $E_{exc}$   
    \\[1ex] \hline 
   $^{\rm 12}$C & $32.8\pm0.8$   &  29.7        &   35.1       
    \\[2ex] \hline 
   $^{16}$O   & $102.6\pm0.6$    &  79.2        &  123.         
    \\[2ex] \hline  
   $^{28}$Si   & $871\pm2$    &   657.       &   970.        
    \\[2ex] \hline  
   $^{40}$Ca  & $2544\pm 7$  &  1490.      &  2730.        
    \\[2ex] \hline 
\end{tabular}

\medskip

\end{center}
\end{table}

\end{document}